# Note on the helicity decomposition of spin and orbital optical currents

Andrea Aiello[1,2] & M V Berry[3]

[1]Max Planck Institute for the Science of Light, Günther-Scharowsky-Strasse 1/Bau24 91058 Erlangen, Germany
[2]Institute for Optics, Information and Photonics, University of Erlangen-Nürnberg, Staudtstrasse 7/B2, 91058 Erlangen, Germany
andrea.aiello@mpl.mpg.de
[3]H H Wills Physics Laboratory, Tyndall Avenue, Bristol BS8 1TL, UK
asymptotico@bristol.ac.uk

**Abstract**

In the helicity representation, the Poynting vector (current) for a monochromatic optical field, when calculated using either the electric or the magnetic field, separates into right-handed and left-handed contributions, with no cross-helicity contributions. Cross-helicity terms do appear in the orbital and spin contributions to the current. But when the electric and magnetic formulas are averaged ('electric-magnetic democracy'), these terms cancel, restoring the separation into right-handed and left-handed currents for orbital and spin separately.

# 1, Introduction

This reports a small development and slight correction to the largely well-understood [1] representation of the current (=time-averaged energy flux =$c^2$× momentum density) by the Poynting vector [2], for a monochromatic optical field in empty space. This can be separated in two ways: into orbital and spin currents, or into positive and negative helicities, and each representation can be expressed in terms of the electric or magnetic field, in turn representable as a superposition of plane waves. Our aim is to clarify the interrelations between these different descriptions, and in particular express the helicity decomposition in a simpler way.

The real time- and space-varying electric field vector $\boldsymbol{E}_{\text{real}}$, with frequency $\omega=ck$, is conveniently expressed in terms of a complex vector $\boldsymbol{E}$ depending only on position $\boldsymbol{r}=(x,y,z)$:

$$\boldsymbol{E}_{\text{real}}(\boldsymbol{r},t) = \text{Re}\left[\boldsymbol{E}(\boldsymbol{r})\exp(-\mathrm{i}\omega t)\right], \tag{1.1}$$

and similarly for the magnetic field $\boldsymbol{H}$. The Poynting vector is

$$\boldsymbol{P} = \tfrac{1}{2}\text{Re}\left[\boldsymbol{E}^* \times \boldsymbol{H}\right], \tag{1.2}$$

where here and hereafter we do not indicate the $\boldsymbol{r}$ dependence explicitly. We are here considering fully three-dimensional fields, unrestricted by paraxiality.

From Maxwell's equations, each field can be expressed in terms of the other. In SI units,

$$\boldsymbol{H} = -\frac{\mathrm{i}}{\omega\mu_0}\nabla\times\boldsymbol{E}, \quad \boldsymbol{E} = \frac{\mathrm{i}}{\omega\varepsilon_0}\nabla\times\boldsymbol{H}. \tag{1.3}$$

giving the equivalent electric and magnetic representations of the current, conveniently written as

$$P = \frac{c^2}{2\omega} P_E = \frac{c^2}{2\omega} P_H, \qquad (1.4)$$

in which

$$P_E = \varepsilon_0 \operatorname{Im}\left[ E^* \times (\nabla \times E) \right] = P_H = \mu_0 \operatorname{Im}\left[ H^* \times (\nabla \times H) \right]. \qquad (1.5)$$

We are here concerned with momentum density, but all results apply, mutatis mutandis, to the angular momentum density, obtained from $P$ by vector-multiplying by $r$, and to characterizations of helicity [3-6].

## 2. Spin and orbital split

By an elementary vector identity, the electric and magnetic currents can be split into parts naturally interpreted as orbital and spin [1]:

$$\begin{aligned}
P_{\text{orb}E} &= \varepsilon_0 \operatorname{Im}\left[ E^* \cdot (\nabla) E \right], & P_{\text{sp}E} &= \frac{\varepsilon_0}{2} \operatorname{Im}\left[ \nabla \times (E^* \times E) \right], \\
P_{\text{orb}H} &= \mu_0 \operatorname{Im}\left[ H^* \cdot (\nabla) H \right], & P_{\text{sp}H} &= \frac{\mu_0}{2} \operatorname{Im}\left[ \nabla \times (H^* \times H) \right],
\end{aligned} \qquad (2.1)$$

where for the orbital currents we use the notation

$$A \cdot (\nabla) B = A_x \nabla B_x + A_y \nabla B_y + A_z \nabla B_z. \qquad (2.2)$$

Although $P_E = P_H$, the separate spin and orbital contributions in $P_E$ and $P_H$ are equal only for paraxial fields. In general, they are different:

$$P_{\text{orb}E} \neq P_{\text{orb}H}, \quad P_{\text{sp}E} \neq P_{\text{sp}H}. \qquad (2.3)$$

This led to the 'electric-magnetic democracy' proposal, in which orbital and spin currents are defined by the average

$$\boldsymbol{P}_{orb} \equiv \tfrac{1}{2}\left(\boldsymbol{P}_{orbE} + \boldsymbol{P}_{orbH}\right), \quad \boldsymbol{P}_{sp} \equiv \tfrac{1}{2}\left(\boldsymbol{P}_{spE} + \boldsymbol{P}_{spH}\right), \qquad (2.4)$$

a stratagem supported by more general considerations [7-10].

## 3. Helicity split

A useful separation of each field is into two components with opposite helicity,

$$\boldsymbol{E} = \boldsymbol{E}_+ + \boldsymbol{E}_-, \quad \boldsymbol{H} = \boldsymbol{H}_+ + \boldsymbol{H}_-, \qquad (3.1)$$

defined by

$$\nabla \times \boldsymbol{E}_\pm = \pm k \boldsymbol{E}_\pm, \quad \nabla \times \boldsymbol{H}_\pm = \pm k \boldsymbol{H}_\pm. \qquad (3.2)$$

These are eigenstates of the momentum projection of the three-dimensional spin operator, through the identity, for any vector $\boldsymbol{A}$,

$$\nabla \times \boldsymbol{A} = \left(\hat{\boldsymbol{p}} \cdot \hat{\boldsymbol{S}}\right)\boldsymbol{A}, \quad \hat{\boldsymbol{p}} = -i\nabla, \quad \hat{\boldsymbol{S}}=\text{vector of spin-1 matrices} \qquad (3.3)$$

(these particular spin matrices, given explicitly in [1], are uniquely defined by the curl). The electric and magnetic helicity components are related by

$$\boldsymbol{H}_\pm = \Box \frac{i}{\mu_0 c} \boldsymbol{E}_\pm, \text{ i.e. } \boldsymbol{H} = -\frac{i}{\mu_0 c}\left(\boldsymbol{E}_+ - \boldsymbol{E}_-\right). \qquad (3.4)$$

The separate contributions correspond to the Riemann-Silberstein vectors [11, 12]

$$\sqrt{\varepsilon_0}\boldsymbol{E}_{\pm} = \pm i\sqrt{\mu_0}\boldsymbol{H}_{\pm} = \tfrac{1}{2}\left(\sqrt{\varepsilon_0}\boldsymbol{E} \pm i\sqrt{\mu_0}\boldsymbol{H}\right). \tag{3.5}$$

When the fields are represented as superpositions of plane waves, the helicities correspond to right- and left-circular polarizations:

$$\boldsymbol{E}_{\boldsymbol{k}\pm} = \boldsymbol{e}_{\pm}\exp(i\boldsymbol{k}\cdot\boldsymbol{r}), \quad \boldsymbol{e}_{\pm} = \tfrac{1}{\sqrt{2}}\left(\boldsymbol{e}_1 \pm i\boldsymbol{e}_2\right)$$
$$\boldsymbol{e}_1 \cdot \boldsymbol{k} = 0, \quad \boldsymbol{e}_2 = \frac{\boldsymbol{k}}{k}\times\boldsymbol{e}_1. \tag{3.6}$$

In singular optics, general helicity eigenstates (i.e. single-helicity superpositions) have the interesting property [12] that their lines of pure circular polarization (C lines [13-15]) for the electric field coincide with those of the magnetic field, and also with the Riemann-Silberstein vortex lines [11].

An advantage of this representation [1, 16] is that the current separates into contributions from the two helicities, i.e.

$$\boldsymbol{P}_E = \boldsymbol{P}_H = \boldsymbol{P}_+ + \boldsymbol{P}_-, \tag{3.7}$$

in which

$$\boldsymbol{P}_{\pm} = \pm\varepsilon_0 k\,\mathrm{Im}\left[\boldsymbol{E}_{\pm}^* \times \boldsymbol{E}_{\pm}\right] = \pm\mu_0 k\,\mathrm{Im}\left[\boldsymbol{H}_{\pm}^* \times \boldsymbol{H}_{\pm}\right]. \tag{3.8}$$

The cross terms, anticipated because $\boldsymbol{P}$ is a quadratic combination of the fields, cancel because of the identity

$$\mathrm{Im}\left[\boldsymbol{E}_+^* \times \boldsymbol{E}_- - \boldsymbol{E}_-^* \times \boldsymbol{E}_+\right] = \mathrm{Im}\left[\boldsymbol{E}_+^* \times \boldsymbol{E}_- + \boldsymbol{E}_+ \times \boldsymbol{E}_-^*\right]$$
$$= \mathrm{Im}\left[2\,\mathrm{Re}\left[\boldsymbol{E}_+^* \times \boldsymbol{E}_-\right]\right] = 0. \tag{3.10}$$

## 4. Combined spin-orbit and helicity separation

Contrary to what has been implied by one of us (sentence after equation (3.25) in [1]), the spin and orbital currents associated with *E* and *H* do not separate into positive- and negative-helicity contributions. In general there are cross terms, i.e.

$$P_{orbE} = P_{orbE+} + P_{orbE-} + P_{orbE+-}, \tag{4.1}$$

where

$$P_{orbE+} = \varepsilon_0 \operatorname{Im}\left[ E_+^* \cdot (\nabla) E_+ \right], \quad P_{orbE-} = \varepsilon_0 \operatorname{Im}\left[ E_-^* \cdot (\nabla) E_- \right],$$
$$P_{orbE+-} = \varepsilon_0 \operatorname{Im}\left[ E_+^* \cdot (\nabla) E_- + E_-^* \cdot (\nabla) E_+ \right] \neq 0. \tag{4.2}$$

Similarly for spin:

$$P_{spE} = P_{spE+} + P_{spE-} + P_{spE+-}, \tag{4.3}$$

where

$$P_{spE+} = \tfrac{1}{2}\varepsilon_0 \nabla \times \operatorname{Im}\left[ E_+^* \times E_+ \right], \quad P_{spE-} = \tfrac{1}{2}\varepsilon_0 \nabla \times \operatorname{Im}\left[ E_-^* \times E_- \right],$$
$$P_{spE+-} = \tfrac{1}{2}\varepsilon_0 \nabla \times \operatorname{Im}\left[ E_+^* \times E_- + E_-^* \times E_+ \right] = \varepsilon_0 \nabla \times \operatorname{Im}\left[ E_+^* \times E_- \right] \neq 0. \tag{4.4}$$

The analogous magnetic contributions are

$$P_{orbH} = P_{orbH+} + P_{orbH-} + P_{orbH+-},$$
$$P_{spH} = P_{spH+} + P_{spH-} + P_{spH+-}, \tag{4.5}$$

related to the corresponding electric currents by (cf. (3.5))

$$P_{orbH+} = P_{orbE+}, \quad P_{orbH-} = P_{orbE-},$$
$$P_{spH+} = P_{spE+}, \quad P_{spH-} = P_{spE-}$$
$$P_{orbH+-} = -P_{orbE+-}, \quad P_{spH+-} = -P_{spE+-}. \tag{4.6}$$

We see that the cross-helicity terms have opposite signs. This leads to the main point we wish to make: the separation into positive and negative helicities of the full current survives the separation into spin and orbital currents if we apply electric-magnetic democracy:

$$\begin{aligned}\boldsymbol{P}_{\text{orb}} &= \tfrac{1}{2}\left(\boldsymbol{P}_{\text{orb}E+} + \boldsymbol{P}_{\text{orb}H+} + \boldsymbol{P}_{\text{orb}E-} + \boldsymbol{P}_{\text{orb}H-}\right),\\ \boldsymbol{P}_{\text{sp}} &= \tfrac{1}{2}\left(\boldsymbol{P}_{\text{sp}E+} + \boldsymbol{P}_{\text{sp}H+} + \boldsymbol{P}_{\text{sp}E-} + \boldsymbol{P}_{\text{sp}H-}\right).\end{aligned} \quad (4.7)$$

Similar results have been obtained before [17, 18], expressed as double Fourier superpositions of plane waves, but our derivation is simpler, and also more general because it allows superpositions that include evanescent waves.

## 5. Example

The simplest illustration of the foregoing general argument is a field composed of a right-circularly polarized plane wave travelling in the $z$ direction and a left-circularly polarized plane wave travelling in the $y$ direction. Choosing units such that $k=1$, with unit direction vectors $\boldsymbol{e}_x$, $\boldsymbol{e}_y$, $\boldsymbol{e}_z$, and ignoring factors $\varepsilon_0$ and $\mu_0$, this field is

$$\begin{aligned}\boldsymbol{E} &= \boldsymbol{E}_+ + \boldsymbol{E}_-,\\ \boldsymbol{E}_+ &= \tfrac{1}{\sqrt{2}}\left(\boldsymbol{e}_x + i\boldsymbol{e}_y\right)\exp(iz),\quad \boldsymbol{E}_- = \tfrac{1}{\sqrt{2}}\left(\boldsymbol{e}_x + i\boldsymbol{e}_z\right)\exp(iy).\end{aligned} \quad (5.1)$$

From (2.1) the orbital and spin currents associated with $\boldsymbol{E}$ and $\boldsymbol{H}$ are

$$\begin{aligned}\boldsymbol{P}_{\text{orb}E} &= \left(\boldsymbol{e}_z + \boldsymbol{e}_y\right)\left(1 + \tfrac{1}{2}\cos(z-y)\right),\quad \boldsymbol{P}_{\text{sp}E} = -\tfrac{1}{2}\left(\boldsymbol{e}_z + \boldsymbol{e}_y\right)\cos(z-y),\\ \boldsymbol{P}_{\text{orb}H} &= \left(\boldsymbol{e}_z + \boldsymbol{e}_y\right)\left(1 - \tfrac{1}{2}\cos(z-y)\right),\quad \boldsymbol{P}_{\text{sp}H} = \tfrac{1}{2}\left(\boldsymbol{e}_z + \boldsymbol{e}_y\right)\cos(z-y),\end{aligned} \quad (5.2)$$

so that

$$P_E = P_H = e_z + e_y, \qquad (5.3)$$

The orbital and spin currents are different for *E* and *H*, and application of electric-magnetic democracy (2.4) gives

$$P_{orb} = \tfrac{1}{2}\left(P_{orbE} + P_{orbH}\right) = e_z + e_y, \quad P_{sp} \equiv \tfrac{1}{2}\left(P_{spE} + P_{spH}\right) = 0. \qquad (5.4)$$

In the helicity representation, the contributions (4.2), (4.4) and (4.5) are

$$\begin{aligned}
&P_{orbE+} = P_{orbH+} = e_z, \quad P_{orbE-} = P_{orbH-} = e_y, \\
&P_{orbE+-} = -P_{orbH+-} = \tfrac{1}{2}\left(e_z + e_y\right)\cos(y-z), \\
&P_{spE+} = P_{spE-} = P_{spH+} = P_{spH-} = 0, \\
&P_{spE+-} = -P_{spH+-} = -\tfrac{1}{2}\left(e_z + e_y\right)\cos(y-z).
\end{aligned} \qquad (5.5)$$

As expected, there are non-zero cross-helicity contributions, but these cancel with the electric-magnetic democracy formula (4.7).